%% file: pra_v11.tex
\pdfoutput=1

\newcommand{\Gal}[1]{\mathbb{F}_{#1}}
\newcommand{\bra}[1]{\ensuremath{\langle#1\vert}}
\newcommand{\ket}[1]{\ensuremath{\vert#1\rangle}}

\documentclass[pra,reprint,showpacs,amsmath,amssymb,superscriptaddress,longbibliography,eqsecnum]{revtex4-1}
\usepackage{graphicx,bm,color,mathptmx}
\usepackage{subfigure}
\usepackage[colorlinks=true,linkcolor=blue,citecolor=blue,urlcolor =blue]{hyperref}

\newcommand{\Tr}{\mathop{\mathrm{Tr}} \nolimits}
\newcommand{\tr}{\mathop{\mathrm{tr}} \nolimits}

\begin{document}

\title{Coarse graining the phase space of $N$ qubits}

\author{Olivia~Di Matteo} 
\affiliation{Department of Physics and Astronomy,
University of Waterloo, Ontario N2L 3G1, Canada} 
\affiliation{Institute for  Quantum Computing, 
University of Waterloo, Ontario N2L 3G1, Canada}

\author{Luis L. S\'{a}nchez-Soto}
 \affiliation{Departamento de  \'Optica, Facultad de F\'{\i}sica, 
 Universidad Complutense, 28040~Madrid, Spain} 
\affiliation{Max-Planck-Institut f\"ur die Physik des Lichts, 
Staudtstra\ss e 2, 91058 Erlangen, Germany}

\author{Gerd Leuchs} 
\affiliation{Max-Planck-Institut f\"ur die Physik des Lichts, 
Staudtstra\ss e 2, 91058 Erlangen, Germany}
\affiliation{Institut f\"ur Optik, Information und Photonik,
  Universit\"{a}t Erlangen-N\"{u}rnberg, Staudtstra{\ss}e 7/B2, 
91058  Erlangen, Germany}

\author{Markus Grassl} 
\affiliation{Max-Planck-Institut f\"ur die Physik des Lichts, 
Staudtstra\ss e 2, 91058 Erlangen, Germany}
\affiliation{Institut f\"ur Optik, Information und Photonik,
  Universit\"{a}t Erlangen-N\"{u}rnberg, Staudtstra{\ss}e 7/B2, 
91058  Erlangen, Germany}

\begin{abstract}
  We develop a systematic coarse graining procedure for systems of $N$
  qubits. We exploit the underlying geometrical structures of the
  associated discrete phase space to produce a coarse-grained version
  with reduced effective size. Our coarse-grained spaces inherit key
  properties of the original ones. In particular, our procedure
  naturally yields a subset of the original measurement operators,
  which can be used to construct a coarse discrete Wigner function. These
  operators also constitute a systematic choice of incomplete
  measurements for the tomographer wishing to probe an intractably
  large system.
\end{abstract}


\maketitle

\section{Introduction}
\label{sec:intro}

Recently, the understanding of many-body quantum systems has dramatically
progressed. Nowadays we are achieving an amazing degree of
control over larger and larger
systems~\cite{Bloch:2008aa,Blatt:2012aa}.  Therefore, verification during
each stage of experimental procedures is of utmost importance;
quantum tomography is the appropriate tool for that purpose.

The goal of quantum tomography is to reconstruct the state of a system
by performing multiple measurements on identically prepared copies of
the system. Once the experimental data are extracted, a numerical
procedure determines which density matrix fits best the
measurements. This estimation can be performed using different
approaches, such as maximum likelihood estimation~\cite{lnp:2004uq},
or Bayesian methods~\cite{Buzek:1998aa,Schack:2001aa,
  Huszar:2012aa,Granade:2016aa}.  However, tomography becomes harder
as we explore more intricate systems. If we look at the simple, yet
illustrative case of $N$ qubits, which will serve as the consistent
thread in this paper, one has to make at least $2^{N}+1$ measurements
in different bases before one can claim to know everything about an
\textit{a priori} unknown system.  With such an exponential scaling in
the number of qubits, it is clear that current methods rapidly become
intractable for present state-of-the-art experiments.

As a result, more sophisticated tomographical techniques are called
for. New protocols try to simplify the process by making an educated
guess about the nature of the state.  Among other assumptions, this
includes rank deficiency~\cite{Gross:2010aa,Cramer:2010aa,
  Flammia:2012aa,Landon-Cardinal:2012aa,Baumgratz:2013aa}, extra
symmetries~\cite{Toth:2010aa,Moroder:2012bs,Klimov:2013aa}, or
Gaussianity~\cite{Rehacek:2009aa}. While all these approaches are
extremely efficient, their pitfall is that when the starting guess is
inaccurate, they produce significant systematic errors.
 
Here, we pursue a different approach, inspired by a notion from statistical
mechanics: coarse graining~\cite{Castiglione:2008aa}.  This operation
transforms a probability density in phase space into a
``coarse-grained'' density that is a piecewise constant function, a
result of density averaging in cells.  This is the chief idea behind
the renormalization group~\cite{White:1992aa}, which allows a
systematic investigation of the changes of a physical system as viewed
at different scales. 

In our case, we consider a system of qubits and look at the associated
phase space, which turns out to be a discrete grid of
$2^{N} \times 2^{N}$ points.  We assign to each suitably defined line
in phase space a specific rank-one projection operator representing a
pure quantum state. For each point of the grid, a suitable
quasi-probability as the Wigner function can be directly computed from
the measurement of the states associated with the lines passing
through that point.  We coarse grain by combining groups of these
lines into thick lines, which we will show to be lines in the phase
space of an effectively smaller system. Our coarse-grained phase
spaces are endowed with many nice properties. 

Most notably, our procedure systematically and naturally reveals a
subset of measurements which one could use to perform incomplete
tomography.  In addition, using the coarse-grained points and lines,
we show that one can define a discrete Wigner function in largely the
same way as it is defined in the original space. When plotted, the
coarse functions resemble smoothed out versions of the originals,
preserving many of their prominent visual features.

\section{Phase space of $N$ qubits}
\label{sec:qpWig}

A qubit is a two-dimensional quantum system, with Hilbert space
isomorphic to $\mathbb{C}^{2}$. It is customary to choose two
normalized orthogonal states, say $\{ | 0 \rangle, | 1 \rangle \}$, as
a computational basis. The unitary matrices
\begin{equation}
  \sigma_{z} = | 0 \rangle \langle 0 | - | 1 \rangle \langle 1 | \, ,
  \qquad \qquad
  \sigma_{x} = | 0 \rangle \langle 1 | + | 1 \rangle \langle 0 | \, ,
  \label{eq:sigmas}
\end{equation}
generate the Pauli group $\mathcal{P}_{1}$, which consists of all the
Pauli matrices plus the identity, with multiplicative factors
$\pm 1, \pm i$~\cite{Chuang:2000fk}.

For $N$ qubits, the corresponding Hilbert space is the tensor product
$\mathbb{C}^{2} \otimes \cdots \otimes \mathbb{C}^{2} =
\mathbb{C}^{2^{N}}$.  A compact way of labeling both states and
elements of the corresponding Pauli group $\mathcal{P}_{N}$ is by
using the finite field $\Gal{2^{N}}$.  In Appendix~\ref{Sec: Galois}
we briefly summarize the basic notions of finite fields needed to
proceed.

Let $| \nu \rangle$, $\nu \in \Gal{2^{N}}$, be an orthonormal basis in
the Hilbert space $\mathbb{C}^{2^{N}}$ (henceforth, field elements
will be denoted by Greek letters).  The elements of the basis can be
labeled by powers of a primitive element $\sigma$ (i.e., a root of an irreducible
primitive polynomial):
$\{|0 \rangle, \, | \sigma \rangle, \ldots, \, |\sigma^{2^N-1} = 1
\rangle \} $. Now the equivalent version of (\ref{eq:sigmas})
is~\cite{Grassl:2003aa,Vourdas:2004aa,Vourdas:2007aa}
\begin{equation}
  Z_{\alpha}  =  \sum_{\nu} \chi ( \alpha \nu  ) \,
  | \nu \rangle \langle \nu | \, ,
  \qquad 
  X_{\beta} = \sum_{\nu}
  |  \nu + \beta \rangle \langle \nu | \, ,
  \label{XZgf} 
\end{equation}
so that
\begin{equation}
  Z_{\alpha} X_{\beta} = \chi ( \alpha \beta ) \, X_{\beta}
  Z_{\alpha}\, ,
\label{commutation_relation}
\end{equation}
which is the discrete counterpart of the Weyl-Heisenberg algebra for
continuous variables~\cite{Binz:2008aa}. Here, the additive character
$\chi $ is defined as $\chi (\alpha ) = \exp [ i \pi \tr ( \alpha ) ]$
and the trace of a field element (we distinguish it from the trace of
an operator by the lower case ``tr'') is defined in Appendix~\ref{Sec:
  Galois}.  Moreover, $Z_{\alpha}$ and $X_{\beta}$ are related through
the finite Fourier transform~\cite{Klimov:2005aa}
\begin{equation}
  \mathcal{F}=\frac{1}{\sqrt{2^{N}}}\sum_{\nu ,\nu^{\prime}}
  \chi (\nu \,\nu^{\prime})\, 
  |\nu \rangle \langle \nu^{\prime}| \, ,
  \label{FTcomp}
\end{equation}
so that $X_{\alpha}=\mathcal{F} \, Z_{\alpha} \,\mathcal{F}^\dag$.

The operators (\ref{XZgf}) generate the Pauli group $\mathcal{P}_{N}$
of $N$ qubits and, with a suitable choice of basis, they can be
factorized into a tensor product of single-qubit Pauli operators.
To this end, it is convenient to consider $\Gal{2^{N}}$ as an
$N$-dimensional linear space over $\mathbb{Z}_{2}$. It is spanned by
an abstract basis $\{ \theta_{1}, \ldots , \theta _{N} \}$, so that
given a field element $\alpha $ the expansion
\begin{equation}
  \alpha = \sum_{i=1}^{N} a_{i} \, \theta _{i} \, , 
  \qquad 
  a_{i}\in \mathbb{Z}_{2} \, ,
  \label{alpha}
\end{equation}
allows us the identification
$\alpha \Leftrightarrow (a_{1}, \ldots, a_{N})$.  The basis
$\{ \theta_{i} \}$ can be chosen to be orthonormal with respect to the
trace operation; i.e.,
$ \tr ( \theta_{i} \, \theta_{j}) = \delta_{ij}$.  This is a self-dual
basis, which always exist for the case of qubits. In this way, we
associate each qubit with a particular element of the self-dual basis:
qubit$_{i} \Leftrightarrow \theta _{i}$. Using this basis, we have the
factorization
\begin{equation}
  Z_{\alpha}  = \sigma_{z}^{a_{1}} \otimes \cdots \otimes
  \sigma_{z}^{a_{N}},
  \qquad 
  X_{\beta} = \sigma_{x}^{b_{1}} \otimes \cdots \otimes
  \sigma_{x}^{b_{N}} \, ,
\end{equation}
where $a_{i} = \tr ( \alpha \theta_{i} )$ and
$b_{i} = \tr ( \beta \theta_{i} ) $ are the corresponding expansion
coefficients for $\alpha$ and $\beta$ in the self-dual basis.

We next recall~\cite{Wootters:2004aa,Gibbons:2004aa} that the grid
defining the phase space for $N$ qubits can be appropriately labeled
by the discrete points $(\alpha, \beta)$, which are precisely the
indices of the operators $Z_{\alpha}$ and $X_{\beta}$: $\alpha$ is the
``horizontal'' axis and $\beta$ the ``vertical'' one.  In this grid we
can introduce the set of displacements
\begin{equation}
  D ( \alpha ,\beta )  =   \Phi (\alpha ,\beta )  \, 
  Z_{\alpha}  X_{\beta} \, ,
  \label{Dop} 
\end{equation}
where $\Phi ( \alpha, \beta)$ is a phase required to avoid plugging
extra factors when acting with $D$. A sensible choice for the case of
qubits is $\Phi^{2} (\alpha , \beta ) = \chi (\alpha \beta)$, which
ensures the Hermiticity of the displacement operators. In addition, we
impose $\Phi (\alpha ,0)=1$ and $\Phi (0, \beta )=1$, which means that
the displacements along the ``position'' axis $\alpha $ and the
``momentum'' axis $\beta $ are not associated with any phase.  These
displacement operators shift phase space points, so the action of
$D(\alpha^{\prime} , \beta^{\prime})$ maps
$(\alpha ,\beta ) {\mapsto} (\alpha +\alpha^{\prime}, \beta +
\beta^{\prime} )$, justifying their designation.  Note that we
still have to fix the sign of the phase $\Phi(\alpha,\beta)$. We
choose the phase as
\begin{equation}
  \Phi (\alpha ,\beta )= i^{\tr(\alpha\beta)}(-1)^{f(\alpha\beta)} \, ,
\end{equation}
where $ f(x)=\sum_{0\le j<i\le m-1} x^{2^i+2^j}$, which ensures that
the operators defined in Eq.~\eqref{eq:line_operators} below are
rank-one projections.

On the phase space grid one can introduce a variety of geometrical
structures with much the same properties as in the continuous
case~\cite{Klimov:2007aa,Klimov:2009aa,Munoz:2012aa}.  The simplest
are the straight lines passing through the origin (also called rays),
with equations
\begin{equation}
  \alpha = 0 ,
  \qquad
  \mathrm{or}
  \qquad
  \beta = \lambda \alpha \, . 
  \label{rays}
\end{equation}
The rays have a very remarkable property: the monomials
$D (\alpha, \beta) $ belonging to the same ray commute, and thus, have
a common system of eigenvectors $\{|\psi_{\nu,\lambda} \rangle \}$,
\begin{equation}
  D (\alpha,  \lambda \alpha ) | \psi_{\nu,\lambda} \rangle =
  \exp(i \xi_{\nu, \lambda}) | \psi_{\nu, \lambda} \rangle ,
  \label{ZXes}
\end{equation}
where $\lambda $ is fixed and $\exp (i \xi_{\nu, \lambda})$ is the
corresponding eigenvalue, so
$| \psi_{\nu, 0} \rangle = | \nu \rangle $ are eigenstates of
$Z_{\alpha}$ (displacement operators labeled by the ray $\beta =0$,
which we take as the horizontal axis). The projection operators
associated with the lines of equal slope are the projections onto
these eigenvactors. Indeed, we have that
\begin{equation}
  \label{eq:unbias}
  |\langle \psi_{\nu, \lambda} | 
  \psi_{\nu^\prime, \lambda^\prime} \rangle |^2 =
  \delta_{\lambda, \lambda^\prime} \delta_{\nu, \nu^\prime}
  + \frac{1}{2^{N}} (1 - \delta_{\lambda, \lambda^\prime}),
\end{equation}
and, in consequence, they are mutually unbiased bases
(MUBs)~\cite{Wootters:1989aa}.

Now suppose for each ray we disregard the origin $(0, 0)$, whose
monomial is the identity operator. This leaves us with $2^N - 1$
commuting operators.  If we then consider the whole bundle of
$2^{N}+1$ rays (which are obtained by varying the ``slope'' $\lambda$
over all of $\Gal{2^N}$), we can construct a complete set of MUB
operators arranged in a $(2^{N}-1) \times (2^{N}+1)$
table~\cite{Bandyopadhyay:2002aa}.

To round up the scenario, we need to represent states in phase
space. The discrete Wigner function~\cite{Bjork:2008aa} is the
appropriate tool. It can be considered as an invertible mapping
\begin{equation}
  W_{\varrho}  (\alpha , \beta  ) = \frac{1}{2^{N}} 
  \Tr [ \varrho \, \Delta  (\alpha , \beta )] \, ,  
  \label{eq:Wigdn}
\end{equation}
so that
\begin{equation}
  \varrho =  \sum_{\alpha ,\beta} 
  \Delta (\alpha  , \beta ) \, W_{\varrho} (\alpha  , \beta ) \, .
\end{equation}
The operational kernel is defined as
\begin{eqnarray}
  \Delta  (\alpha , \beta )  =  \frac{1}{2^{N}} 
  \sum_{\alpha^{\prime} ,\beta^{\prime}}  
  \chi (\alpha \alpha^{\prime} - \beta \beta^{\prime} ) \,
  D(\alpha^{\prime} ,\beta^{\prime} ) \, ,
\end{eqnarray}
which, in view of equation~(\ref{FTcomp}), can be interpreted as a
double Fourier transform of $D (\alpha ,\beta )$. One can check that
this kernel has all the good properties~\cite{Stratonovich:1956aa}: it
is Hermitian, normalized and covariant under the Pauli group. As a
result, for each point on the grid, the
corresponding value of the Wigner function can be computed from the probabilities of
measuring the pure states associated with the lines passing through
that point.

\begin{figure*}[t]
  \includegraphics[width=1.75\columnwidth]{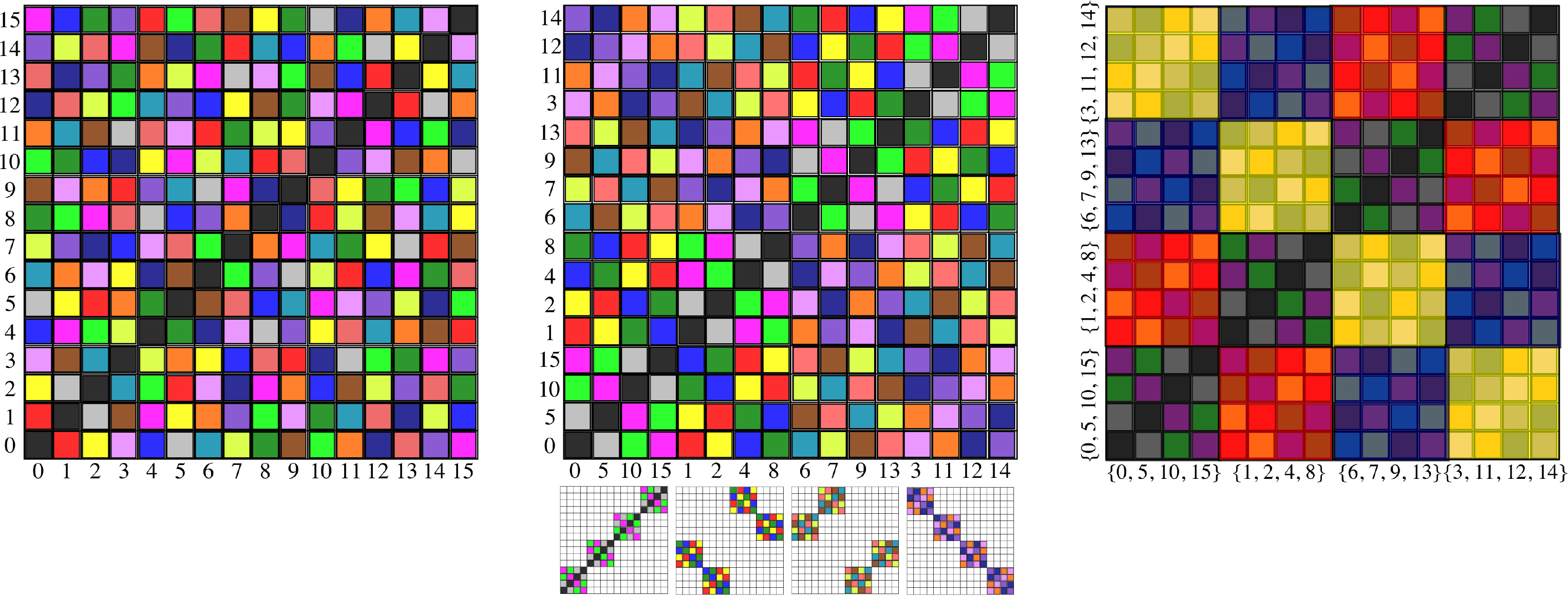}
  \caption{Graphical sketch of coarse graining. Here we consider
    dimension 16, and its diagonal ray, $\beta = \alpha$.  The first
    panel plots all the lines of the form $\beta = \alpha + \gamma$,
    parametrized by the shift $\gamma$. Points on the same line have
    the same colour.  Axis labels correspond to powers of the
    primitive element of $\Gal{16}$, with the convention that
    $\sigma^0$ is denoted by 0 and $\sigma^{15} = 1$. The middle panel shows the
    original grid with the axis labels permuted such that coset
    elements are grouped together. We can see that this leads to
    distinct $4\times 4$ blocks containing points of exactly four
    different colours. These are shown expanded out in the small,
    lower four grids. One notices that these ``coarse'' blocks form
    the diagonal ray and all its translates in dimension $4$, which we
    show superimposed in the last panel.}
  \label{colourfulpanels}
\end{figure*}

\section{Coarse graining}

As heralded in the Introduction, our goal is to tailor a procedure
that allows us to coarse grain the phase space of a multiqubit system;
i.e., to break it down into simpler sub-components.

To this end, we consider the number $N$ of qubits to be composite,
i.e. $N=mn$.  Let $\{ \mu_{0}, \ldots, \mu_{n-1}\}$ be a basis of
$\Gal{2^{mn}}$ with respect to $\Gal{2^{m}}$.  We define
\begin{equation}
  \mathfrak{C}_{0} = \left\{ \sum_{j = 1}^{n-1}  \tau_{j} \mu_{j} \enskip | 
    \enskip \tau_j \in \Gal{2^{m}} \right \} \, ,
\end{equation}
i.e., the subspace made of linear combinations of basis elements
$\mu_{1}, \ldots, \mu_{n-1}$ with coefficients in the base field
$\Gal{2^m}$.  We can use this set $\mathfrak{C}_0$, which we
henceforth refer to as the initial coset, to decompose the field
$\Gal{2^{mn}}$ into cosets:
\begin{equation}
  \mathfrak{C}_{\tau} = \tau \mu_{0} + \mathfrak{C}_{0}, \qquad 
  \tau \in \Gal{2^{m}} \, .
\end{equation}
The coarse-grained space will be labeled according to these cosets.

We can imagine the process of coarse graining as partitioning the grid
$\Gal{2^{mn}} \times \Gal{2^{mn}}$ in such a way that we superimpose a
grid of size $2^{m} \times 2^{m}$ on top, with each
superimposed point indexed by cosets rather than field elements in the
original grid. Each point in the coarse grid then contains a sub-grid
the same size as $\Gal{2^{m}}^{n-1} \times \Gal{2^{m}}^{n-1}$.  To
provide some intuition for this, we show a visual example of this
process in action in Fig.~\ref{colourfulpanels}.

Our procedure for coarse-graining the grid arises naturally from
consideration of the line structure of phase space.  We
will use the \emph{thin} lines in $\Gal{2^{mn}}$ to create
\emph{thick} lines in the coarse phase space, by grouping together
lines having the same slope, and with intercepts in the same coset. We write thin
lines in the big field $\Gal{2^{mn}}$ as
$\ket{\ell^{(\lambda)}_\gamma}$, where $\lambda$ is the slope, and
$\gamma$ is the intercept. A large, coarse-grained line is denoted as
$\ket{L^{(\lambda)}_{\mathfrak{C}_{\tau}}}$, where now the intercept
is a whole coset.

To each line in the fine-grained phase space we can assign a projector
$\ket{\ell^{(\lambda)}_\gamma}\bra{\ell^{(\lambda)}_\gamma}$,
constructed as a linear combination of the displacement operators.  We
choose as our convention for the rays $(\gamma = 0)$ the all-positive
sum
\begin{equation}
  \ket{\ell^{(\lambda)}_0}\bra{\ell^{(\lambda)}_0}
  = \frac{1}{2^{mn}} \sum_{\alpha} D(\alpha, \lambda \alpha).
  \label{eq:line_operators}
\end{equation}
These lines are eigenstates with eigenvalue $+1$ for all displacement
operators in the sum. Projectors with nonzero intercepts are obtained
by conjugating that of the ray with an appropriate displacement
operator. 

The coarse lines are produced by grouping together lines with
intercepts in the same coset:
\begin{equation}
  \ket{L^{(\lambda)}_{\mathfrak{C}_{\tau}}}\bra{L^{(\lambda)}_{\mathfrak{C}_{\tau}}} 
  = \sum_{\gamma \in \mathfrak{C}_{\tau}} 
  \ket{\ell^{(\lambda)}_\gamma}\bra{\ell^{(\lambda)}_\gamma} \, . 
  \label{coarse_untransformed}
\end{equation}
The possible choices of slope for these lines will be limited to
elements of the subfield $\Gal{2^m}$, as these have natural analogues
between the two fields.

As discussed in more detail in Appendix~\ref{Sec: Lineop}, the coarse
rays of Eq.~(\ref{coarse_untransformed}) can be simplified and
rewritten as the sum of displacement operators
\begin{equation}
  \ket{{L}^{(\lambda)}_{\mathfrak{C}_0}}
  \bra{{L}^{(\lambda)}_{\mathfrak{C}_0}} = \frac{1}{2^{mn}}
  \sum_{\lambda} \left [ \sum_{\gamma \in \mathfrak{C}_0} 
    \chi  (\gamma \alpha) \right ] D(\alpha, \lambda \alpha)  \, .
  \label{master_eq}
\end{equation}

One can check here that the inner sum over the elements of
$\mathfrak{C}_0$ will cause some of the displacement operators to
vanish. The sum in brackets in Eq.~\eqref{master_eq} is either zero or
a positive constant.  Hence, the projection associated to the thick
lines are a sum over a subset of the displacement operators associated
with the thin lines. This leads us to the key idea of our work: rather than
measuring all the displacement operators, we measure only those which
are present in the rays of the coarse-grained space.

We note here that the choice of $\mathfrak{C}_0$ is not unique, and
will ultimately determine the resultant set of displacement
operators. For example, a special case occurs when the dimension of
the system is square. In this case, we can consider the relationship
between the fields as a quadratic field extension, i.e. when $n = 2$.
In this case we can partition $\Gal{2^{2m}}$ into
$\Gal{2^m} \times \Gal{2^m}$. We can then choose the initial coset as
the copy of the subfield $\Gal{2^m} \subset \Gal{2^{2m}}$:
\begin{equation}
  \mathfrak{C}_0 = \{ \sigma^{i \left(2^m + 1 \right)}, 
  \enskip i = 0, \ldots, 2^m - 1 \} \, ,
  \label{subfield_in_big_field}
\end{equation}
where $\sigma$ is a primitive element of $\Gal{2^{mn}}$ and we use the
notation $\sigma^0$ for 0. The subsequent cosets are
obtained additively from this subfield using the representatives
$\tau_{i} = \sigma^{2^m \left( i - 1 \right) + i}$.

Finally, the coarse-grained phase space inherits a coarse-grained
Wigner function. A coarse kernel can be constructed by grouping
together kernel operators from the same coset, i.e.
\begin{equation}
  \mathfrak{D}  (\mathfrak{C}_{\tau}, \mathfrak{C}_{\xi} ) =
  \sum_{\alpha \in \mathfrak{C}_{\tau}}
  \sum_{\beta \in   \mathfrak{C}_{\xi}} 
  \Delta(\alpha, \beta) \, .
\end{equation}
Desired properties of a Wigner function all follow from the original
kernel. As was the case with the displacement operators, differing
choices of the subset $\mathfrak{C}_0$ will lead to differing Wigner
functions.

\section{Examples}

\begin{figure}[b]
  \centering
  \includegraphics[width=\columnwidth]{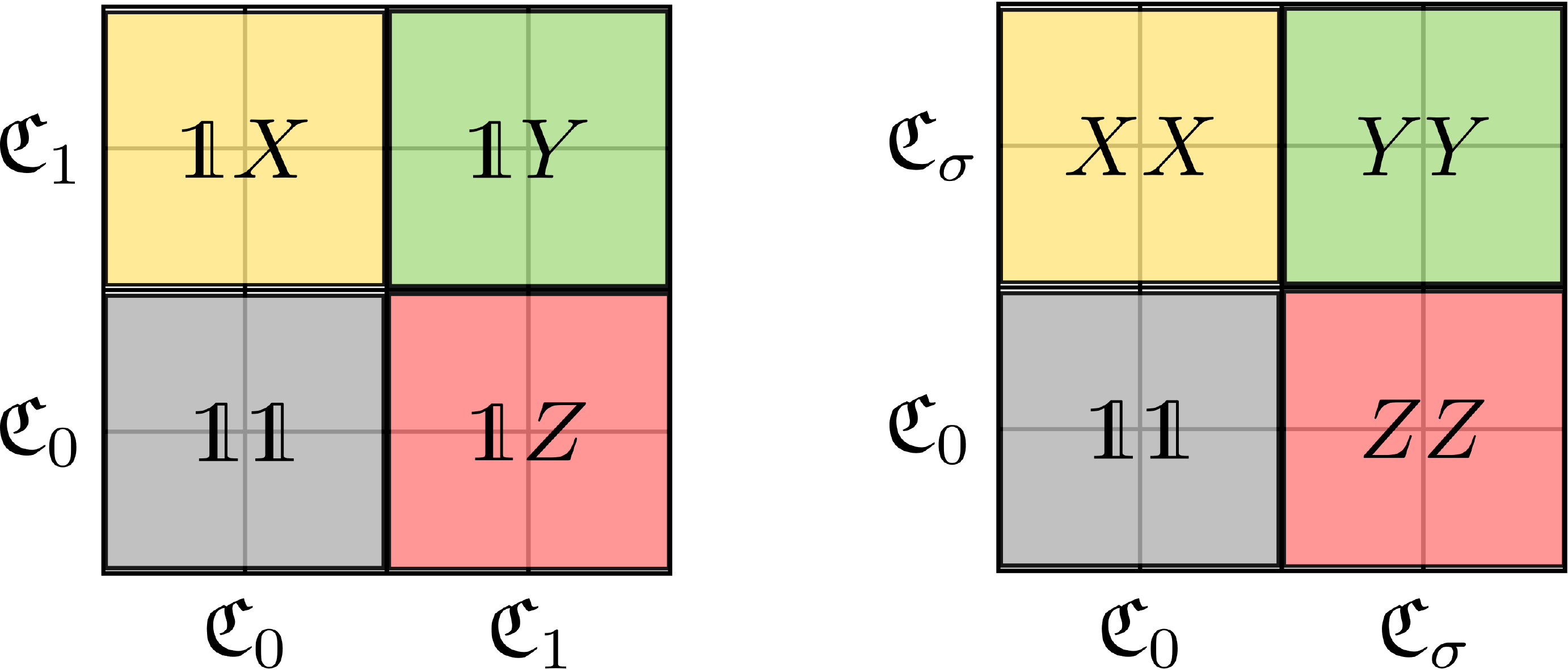}
  \caption{Resultant operators from coarse-graining a dimension 4 system down to dimension
    $2$. Colours are indicative of particular coarse rays. The left image coarse grains by taking
    $\mathfrak{C}_0 = \{ 0, \sigma \}$, whereas the lower image uses
    the subfield $\mathfrak{C}_0 = \{0, 1 \} $.}
  \label{dim4_plane_rays}
\end{figure}

We illustrate the previous ideas with some relevant examples. 
We have written a Python software package capable of generating all
the following results, which we make available online~\cite{Balthasar:2016}

The first nontrivial instance we can have is the case of two qubits, so
dimension $4$.  Using the irreducible
primitive polynomial $x^2 + x + 1 = 0$, we have that
$ \Gal{4} = \{0, 1, \sigma, \sigma^2 = \sigma + 1 \}$.  The self-dual basis is
$\{\sigma, \sigma + 1 \}$, and we use it to produce the displacement
operators.

Another basis for $\Gal{4} / \Gal{2}$ is $\{1, \sigma \}$. Taking all
scalar multiples of $\mu_1 = \sigma$ from the prime field gives us
$\mathfrak{C}_{0} = \{0, \sigma \}$. We then obtain
$\mathfrak{C}_{1} = 1 + \mathfrak{C}_{0} = \{1, \sigma^2 \}$.  For
each ray, we can list the operators which survive in the inner sum
over $\mathfrak{C}_0$ in Eq.~(\ref{master_eq}). Moreover, we can label
the points of the coarse-grained grids by those displacement
operators. Disregarding the identity operator, the resulting set
$\{\openone X, \openone Z, \openone Y \}$ constitutes the appropriate
measurements to be performed to determine which coarse-grained line
they are in. They are essentially Pauli measurements on one of the two
qubits in the system.

Alternatively, the dimension is a square, so we can choose as our initial
coset the subfield $\Gal{2}$: $\mathfrak{C}_{0} = \{0, 1\}$.  This
yields the second coset
$\mathfrak{C}_{\sigma} = \{ \sigma, \sigma^2 \}$. We once again
compute the surviving operators using Eq.~\eqref{master_eq}.  The
final result now is $\{X X, Y Y, Z Z\}$. Here, we see that we are
making a measurement with the same Pauli operator on both
qubits, thereby capturing the full correlations between the two
qubits.  Figure~\ref{dim4_plane_rays} shows both partitioning methods
side by side.

\begin{figure}
  \centering
  \includegraphics[width=0.9\columnwidth]{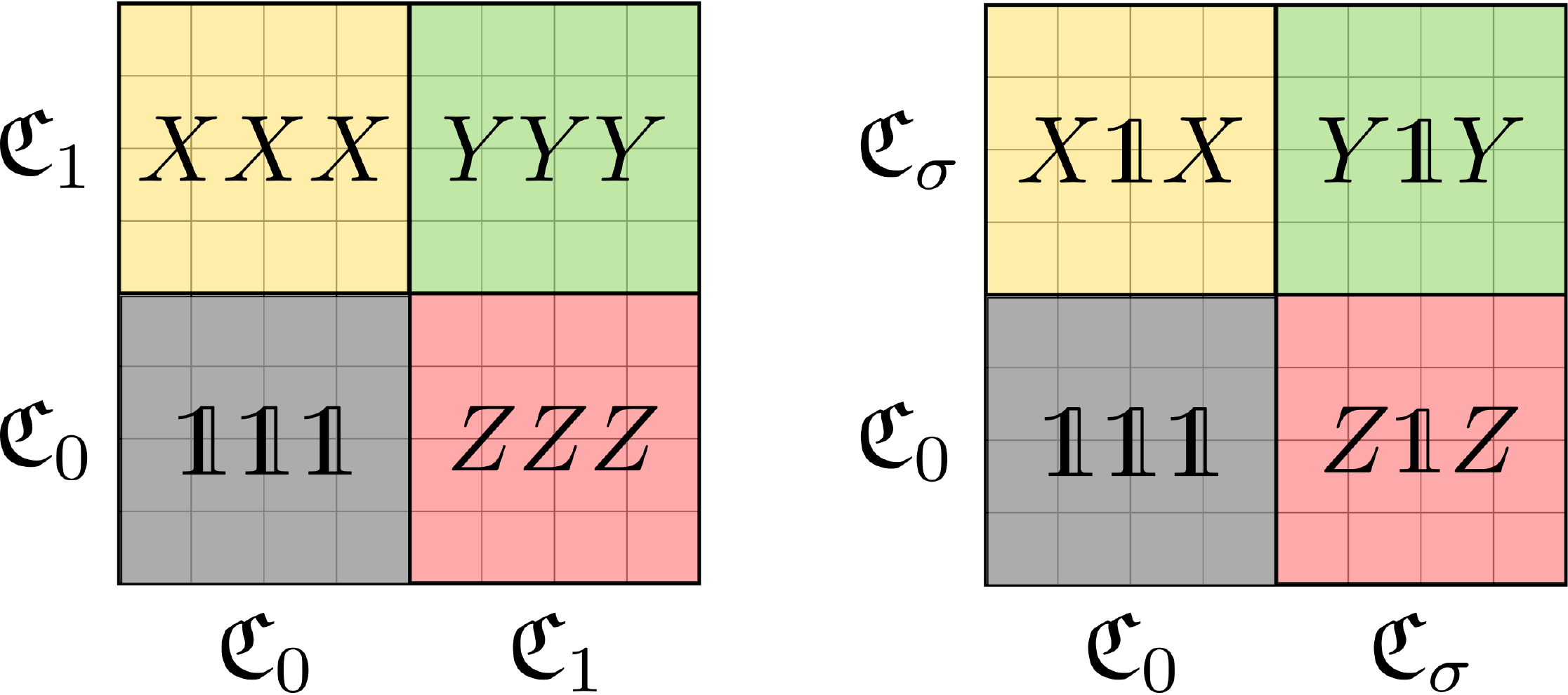}
  \caption{Resultant operators from coarse-graining a dimension $8$
    system down to dimension $2$. (Left panel) Coarse graining using
    the basis $\{1, \sigma, \sigma^2 \}$.  The resultant measurements
    are unitarily equivalent to a case where two of the qubits remain
    untouched. (Right panel) Resultant operators when the
    coarse-graining uses the initial basis
    $\{ \sigma, \sigma^4, \sigma^5 \}$. Here we obtain the interesting
    result that all resultant operators commute.}
  \label{dim8_dim2coarse}
\end{figure}
\begin{figure}[b]
  \centering \includegraphics[width=0.95\columnwidth]{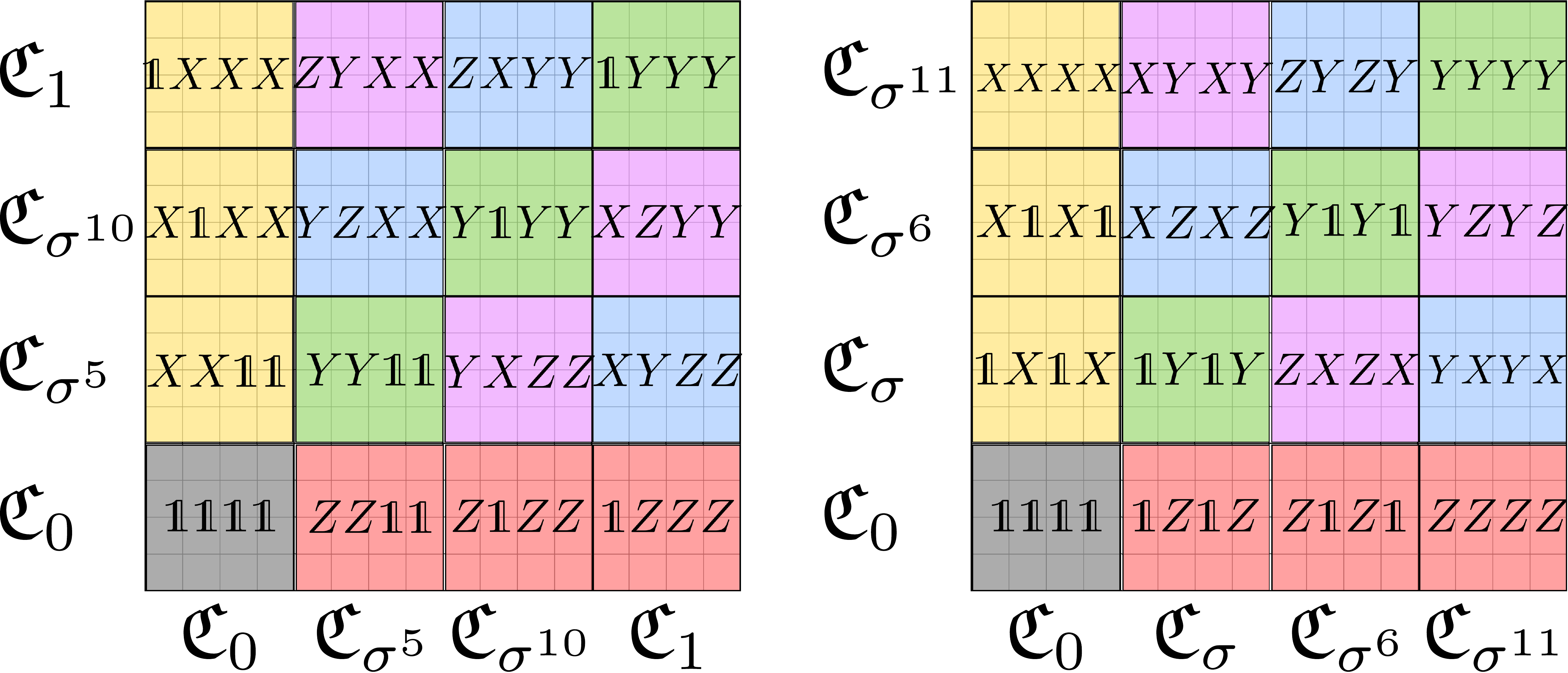}
  \caption{Resultant operators from coarse-graining a dimension $16$
    system down to dimension $4$. The left panel contains the surviving operators
    from the general basis method, the right panel from choosing the
    subfield as $\mathfrak{C}_0$.  The cosets are listed in
    Eqs.~(\ref{dim16cosets_general}) and (\ref{dim16cosets_subfield})
    respectively.  In the case of the left panel, these operators are
    unitarily equivalent to a set where two qubits are untouched and
    the 2-qubit MUB operators are applied to the rest. The right panel
    has no such transformation.}
  \label{dim16_dim4_general}
\end{figure}

\begin{figure*}
  \centering
  \subfigure{\includegraphics[width=0.65\columnwidth]{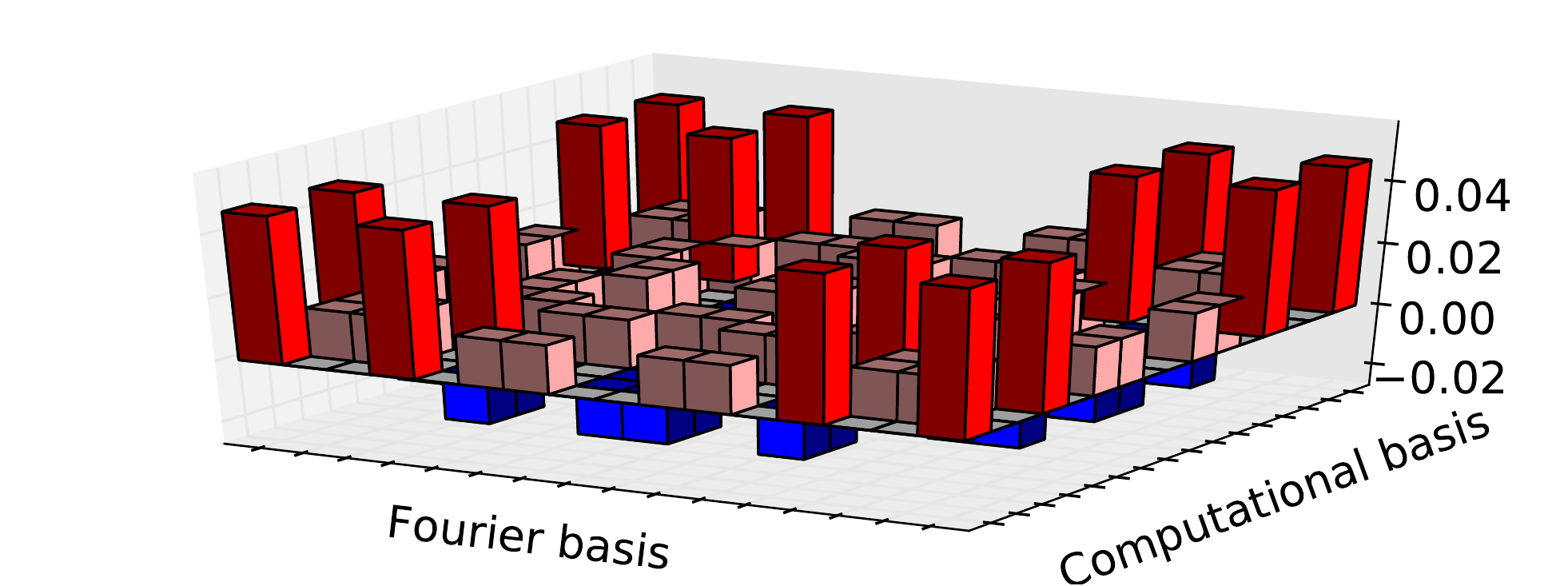}}
  \subfigure{\includegraphics[width=0.65\columnwidth]{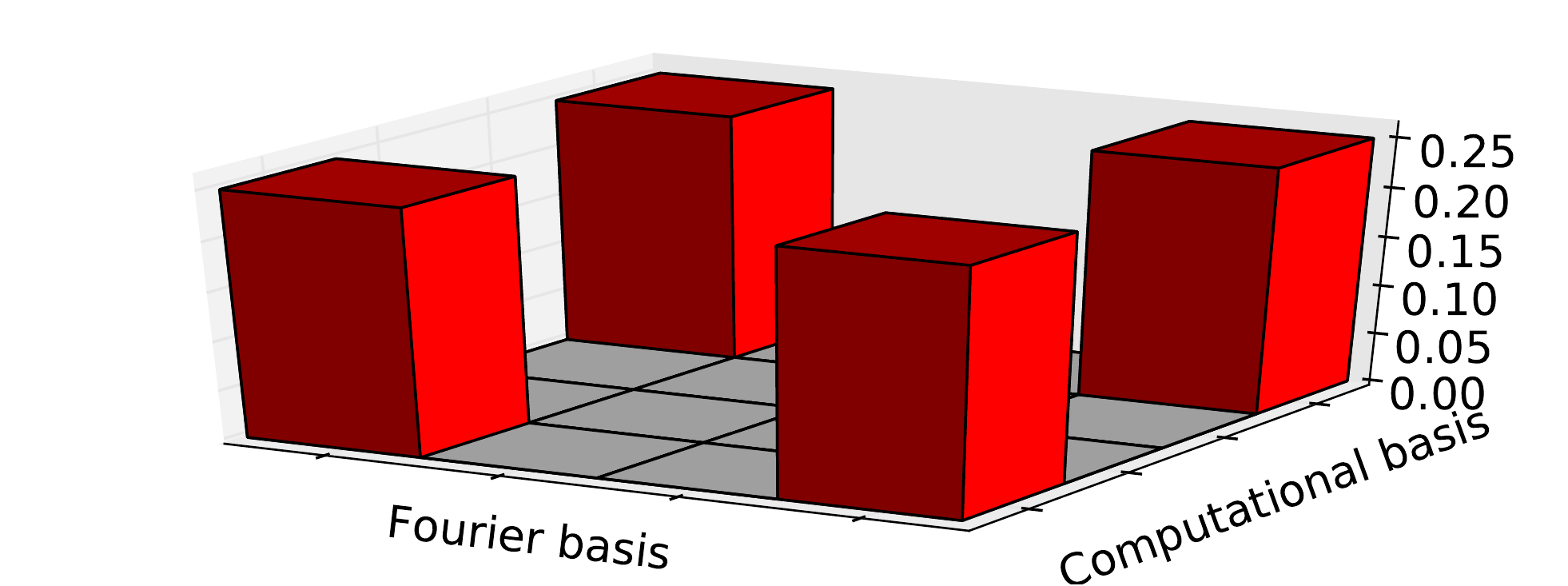}}
  \subfigure{\includegraphics[width=0.65\columnwidth]{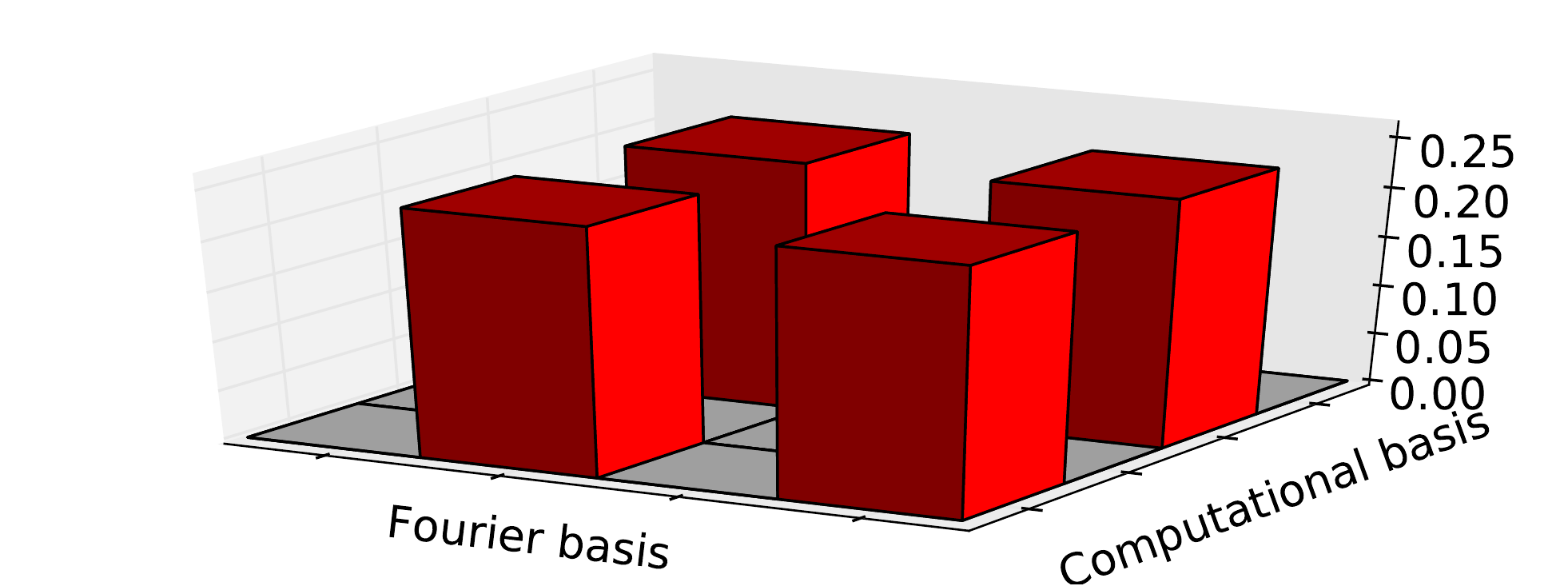}}
  \subfigure{\includegraphics[width=0.65\columnwidth]{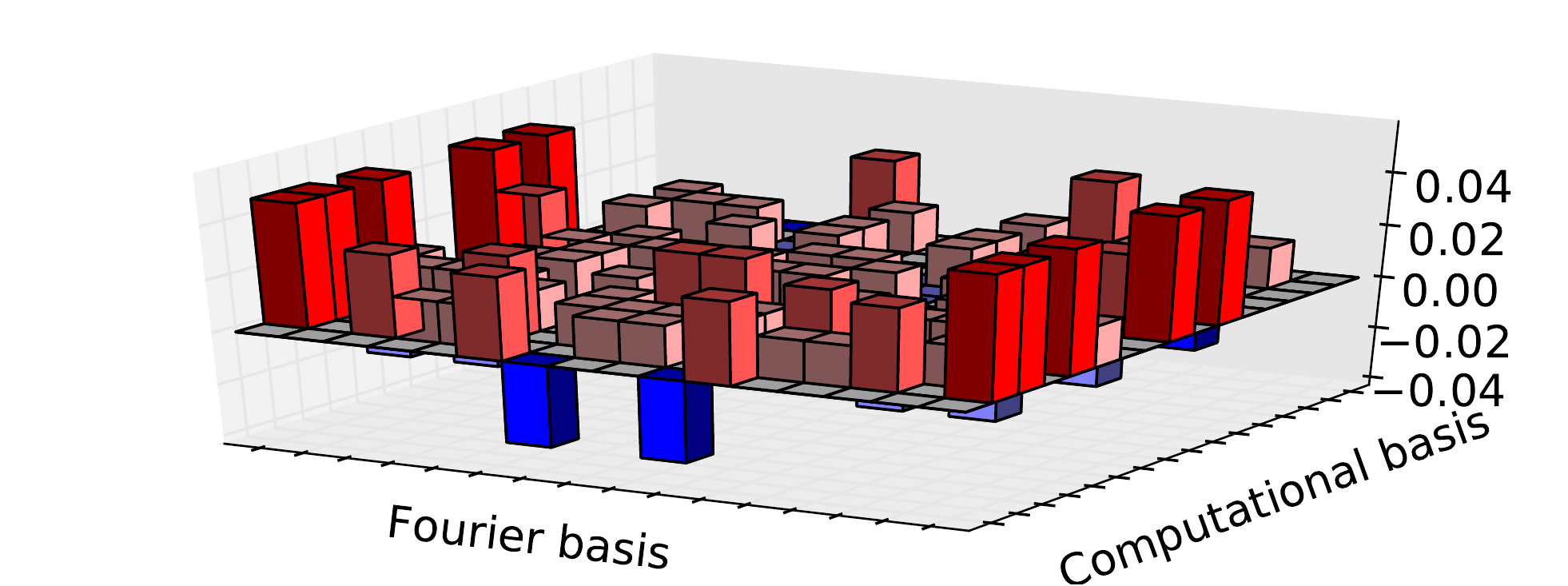}}
  \subfigure{\includegraphics[width=0.65\columnwidth]{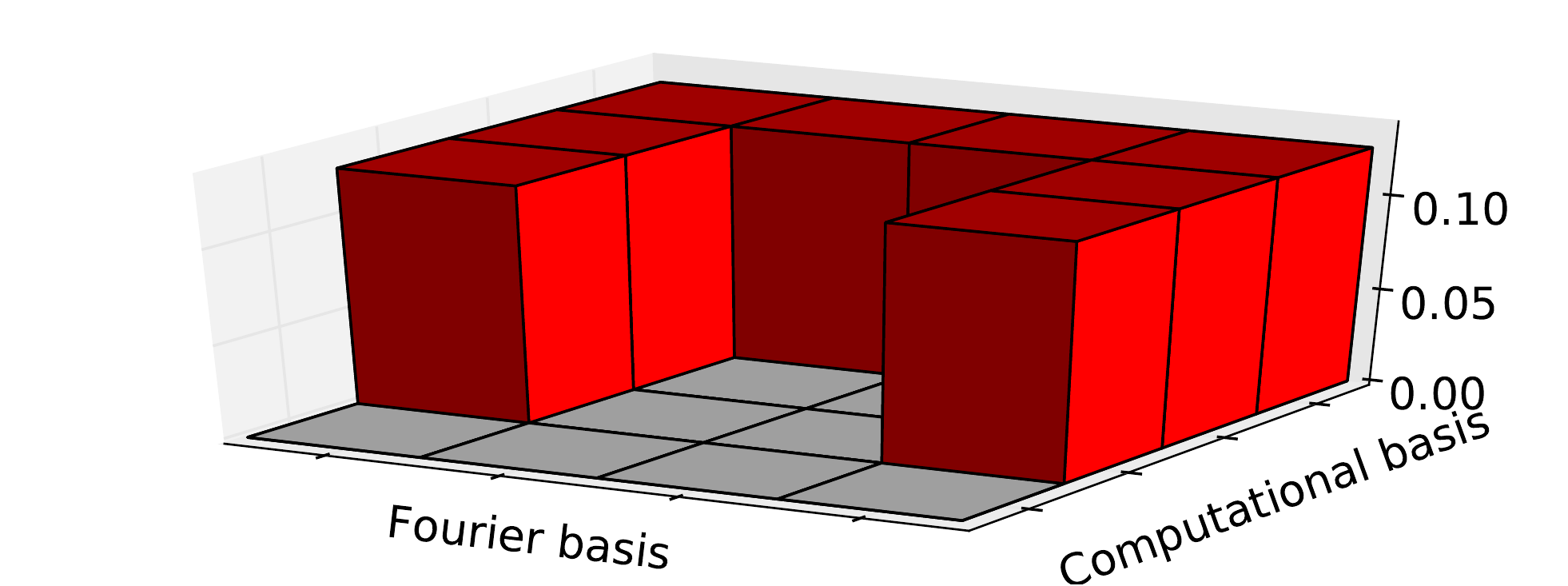}}
  \subfigure{\includegraphics[width=0.65\columnwidth]{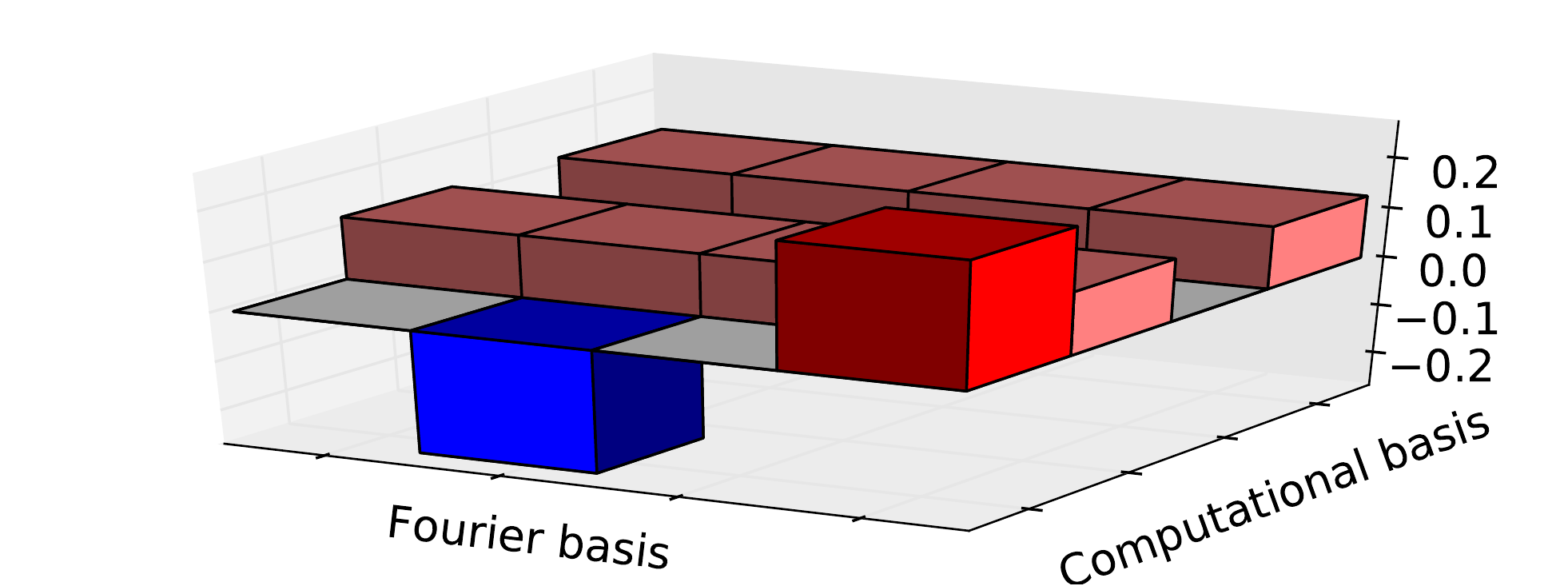}}
  \caption{(Top) Coarse-grained Wigner function for the state
    $\frac{1}{2}(\ket{00} + \ket{11}) \otimes (\ket{00} + \ket{11})$.
    (Left) The original Wigner function in dimension 16.  The $x$-axis
    represents the computational basis, in the standard ordering
    $|0000\rangle, |0001 \rangle, |0010 \rangle$, etc. The Fourier
    basis, as defined via Eq.~\eqref{FTcomp}, is on the $y$-axis and
    is similarly ordered.  (Centre) Coarse graining over $\Gal{4}$
    with the polynomial basis $\{1, \sigma \}$. Here the axes are not
    labeled by single states, but rather by a set of states associated
    with each coset.  (Right) Coarse graining with the subfield as the
    initial coset.  (Bottom) The same coarse graining procedure as
    above, but applied to the state
    $\frac{1}{2}(\ket{0001} + \ket{0010} + \ket{0100} + \ket{1000})$.}
  \label{double_bell_wf}
\end{figure*}

Our next example is the case of dimension 8. We choose $\sigma$ a root of the
  irreducible primitive polynomial $x^3 + x + 1 = 0$, and obtain a
  self-dual basis $\{\sigma^3, \sigma^5, \sigma^6 \}$. An obvious
choice for a basis of $\Gal{8} / \Gal{2}$ is a polynomial basis
$\{1, \sigma, \sigma^2 \}$. To construct $\mathfrak{C}_0$, we
must take all possible linear combinations of $\sigma$ and $\sigma^2$
with coefficients in $\Gal{2}$. This produces
\begin{equation}
  \mathfrak{C}_0 =   \{0, \sigma, \sigma^2, \sigma^4 \}.
\end{equation}
We obtain the second coset by adding the remaining subfield element
$1$ to $\mathfrak{C}_0$:
\begin{equation}
  \mathfrak{C}_1 = \{1, \sigma^3, \sigma^5, \sigma^6 \}.
\end{equation}
The traces of all elements in $\mathfrak{C}_0$ are $0$, and the
  traces for all elements in $\mathfrak{C}_1$ are $1$.  The surviving
four operators are shown in Fig.~\ref{dim8_dim2coarse}.

Using a Clifford transformation, we can ``trace out'' two of the
qubits.  The sequence of CNOT gates: {CNOT}$_{12}$ -- {CNOT}$_{13}$ --
{CNOT}$_{21}$ -- {CNOT}$_{31}$ transforms the set into
$\{X\openone \openone, Z \openone \openone, Y \openone \openone\}$, so
we see that this partitioning is, after a global change of basis,
equivalent to measuring each Pauli on only a single qubit.

If we choose instead the basis $\{\sigma, \sigma^4, \sigma^5\}$ to
build our cosets, we get a more interesting result:
\begin{equation}
  \mathfrak{C}_0 =  \{0, 1, \sigma^4, \sigma^5 \}, 
  \qquad
  \mathfrak{C}_\sigma = \{\sigma, \sigma^2, \sigma^3, \sigma^6 \}.
\end{equation}
The operators that survive have the form $Z_\alpha X_\beta$,
$\alpha, \beta \in \{0, \sigma^4 \}$, yielding the operators in
Fig.~\ref{dim8_dim2coarse}, which all commute.  In this case, we are
already ignoring one of the three qubits. However, it is not possible
to find a Clifford which will trace out a remaining one as was the
case with the polynomial basis case.  So, in a sense, using this
partitioning we are ignoring fewer qubits than before.

Dimension $16$ is perhaps the first really interesting case. First of
all, we can consider it in two ways: $m = 1, n = 4$, or
$ m = 2, n = 2$. Essentially, to do the partitioning, we can look at
$\Gal{16}$ as a quartic extension over $\Gal{2}$, or a quadratic
extension over $\Gal{4}$. We consider the quadratic case, so we can
coarse grain in two ways.  We work
with $\Gal{16}$ as constructed by the irreducible primitive polynomial
$x^4 + x + 1$ over $\Gal{2}$, and $x^2 + x + \sigma^\prime$ over
$\Gal{4}$ where we denote a primitive element of $\Gal{4}$ as
$\sigma^\prime$. We know from Eq.~(\ref{subfield_in_big_field}) that
$\sigma^\prime = \sigma^5$, where $\sigma$ is the primitive element in
$\Gal{16}$. Then $\Gal{4}$ in $\Gal{16}$ can be written as
$\{0, \sigma^5, \sigma^{10}, \sigma^{15} = 1 \}$.

For the general case, we choose the basis $\{1, \sigma \}$. Taking all
$\Gal{4}$-multiples of $\sigma$, we obtain
$\mathfrak{C}_0 = \{0, \sigma, \sigma^6, \sigma^{11} \}$. The full set
of cosets is:
\begin{equation}
  \begin{array}{ll}
    \mathfrak{C}_0  = \{0, \sigma, \sigma^6, \sigma^{11} \},  &
    \mathfrak{C}_{\sigma^5} = \{\sigma^5, \sigma^2, \sigma^9, \sigma^3 \},
    \\
    \mathfrak{C}_{\sigma^{10}} = \{ \sigma^{10}, \sigma^8, \sigma^7,
    \sigma^{14} \},  &
   \mathfrak{C}_1 = \{1, \sigma^4, \sigma^{13}, \sigma^{12} \}. 
  \end{array}
  \label{dim16cosets_general} 
\end{equation}

Proceeding in the standard way, and taking into account that a
self-dual basis is
$\{ \sigma^3, \sigma^7, \sigma^{12}, \sigma^{13} \}$, we obtain the
operators in Fig. \ref{dim16_dim4_general}.  What is (un)interesting
about these operators is that we can transform them all into operators
which completely ignore two of the qubits. In particular, consider the
following sequence of operations: {CNOT}$_{43}$ -- {CNOT}$_{32}$ --
{CNOT}$_{31}$ -- {CNOT}$_{14}$ -- {CNOT}$_{24}$. Application of this
to the operators of the first panel of Fig.~\ref{dim16_dim4_general}
yields a new set of operators where the last two qubits contain only
$\openone$, and the first two qubits contain the full set of MUB
operators on two qubits.

Alternatively, we can choose our initial coset as the subfield, and
the coset representatives as $\tau_i = \sigma^{4(i-1) + i}$. We obtain
the cosets
\begin{equation}
  \begin{array}{ll}
    \mathfrak{C}_0 =  \{0, 1, \sigma^5, \sigma^{10} \}, &
   \mathfrak{C}_{\sigma} = \{\sigma, \sigma^4, \sigma^{2}, \sigma^{8} \},   \\ 
    \mathfrak{C}_{\sigma^{6}} = 
    \{ \sigma^{6}, \sigma^{13}, \sigma^9, \sigma^{7} \},  & 
    \mathfrak{C}_{\sigma^{11}}  = 
 \{\sigma^{11}, \sigma^{12}, \sigma^{3},\sigma^{14} \}.
  \end{array}
  \label{dim16cosets_subfield}
\end{equation} 
Using Eq.~(\ref{master_eq}) we get the table shown in the right panel
of Fig.~\ref{dim16_dim4_general}. Unlike in the previous case, there
is no transformation which will lead to us `tracing out' two of the
qubits.  However, we can bring these operators into a more basic form
by applying the sequence {CNOT}$_{13}$ -- {CNOT}$_{24}$.  The resultant operators have the 
property that on the first two
qubits, we only have $X$, and on the last two qubits only $Z$, so that they all  commute.

To conclude, we present some of the coarse-grained Wigner functions we
obtain using our method. Those in dimensions 4 and 8 are somewhat
trivial, so we focus on dimension 16. Wigner functions for the states
$\frac{1}{2}(\ket{00} + \ket{11})(\ket{00} + \ket{11})$ and
$\frac{1}{2}(\ket{0001} + \ket{0010} + \ket{0100} + \ket{1000})$ are
presented in Fig.~\ref{double_bell_wf}.

Recall in Section \ref{sec:qpWig} that we could associate the elements
of $\Gal{2^N}$ with a basis in our Hilbert space. Then, in the coarse
Wigner functions, when we group the field elements into cosets, we can
consider this also as grouping together the associated basis
states. Hence, the probabilities in these Wigner functions become
distributed over the cosets which contain the constituent basis states
of our target state. As a result, the coarse Wigner functions resemble
`smoother' versions of the original one to varying degrees.

\section{Conclusions}

Compared to the continuous Wigner function, the discrete Wigner
function is an adolescent formulation, slowly developing into adult
maturity.  Discrete phase space imposes several new
challenges, which leads to an intricate mapping of the Wigner
function. 

Our coarse graining procedure shows a way to facilitate our
understanding when the number of qubits is high.  While it is always
possible to ignore part of the system and to determine the full Wigner
function of the resulting reduced density matrix, our approach allows
more choices regarding which information of the whole system is measured.  In
another extremal case, the coarse-grained Wigner function is completely
determined by a set of commuting operators that can be measured
simultaneously.

However, several open
questions remain. An obvious next step would be to extend the coarse
graining procedure to multi-qudit systems. Furthermore, knowing the
coarse-grained function, does there exist another subset of
measurements which will allow us to zoom in on specific areas of it
and gain more information? A logical first choice would be to extend
the set of measurements such that they include all operators 
that correspond to slopes in the subfield. For example, in the
dimension $16$ case, we would measure all operators for the rays
$\alpha = 0$ and $\beta = \lambda \alpha, \lambda \in 
\{0, \sigma^5, \sigma^{10}, \sigma^{15}\}$, rather than just three
from each.  This strategy would
allow us to optimize measurements in a very subtle way. Work along
these lines is in progress.

\section{Acknowledgments}

O.D.M. is funded by Canada's NSERC. She is also grateful for
hospitality at the MPL. IQC is supported in part by the
Government of Canada and the Province of
Ontario. L.L.S.S. acknowledges financial support from the Spanish
MINECO (Grant No. FIS2015-67963-P).

\appendix

\section{Finite fields}
\label{Sec: Galois}

In this appendix we briefly recall some background needed for this
paper.  The reader interested in more mathematical details is
referred, e.g., to the excellent monograph by Lidl and
Niederreiter~\cite{Lidl:1986fk}.

A commutative ring is a nonempty set $R$ with two binary
operations, called addition and multiplication, such that it is an
Abelian group with respect to addition, and the multiplication is
associative. The most typical example is the ring of integers
$\mathbb{Z}$, with the standard sum and multiplication. On the other
hand, the simplest example of a finite ring is the set $\mathbb{Z}_n$
of integers modulo $n$, which has exactly $n$ elements.

A field $F$ is a commutative ring with division, i.e., such that 0
does not equal 1 and all elements of $F$ except 0 have a
multiplicative inverse (note that 0 and 1 here stand for the identity
elements for the addition and multiplication, respectively, which may
differ from the familiar real numbers 0 and 1). Elements of a field
form Abelian groups with respect to addition and multiplication (in
this latter case, the zero element is excluded). Note that the finite
ring $\mathbb{Z}_d$ is a field if and only if $d$ is a prime number.

The characteristic of a finite field is the smallest positive
integer $d$ such that
\begin{equation}
   \underbrace{1 + 1 + \ldots + 1}_{\mbox{\scriptsize $d$ times}}=0
\end{equation}
and it is always a prime number. Any finite field contains a prime
subfield $\mathbb{Z}_d$ and has $d^n$ elements, where $n$ is a natural
number. Moreover, the finite field containing $d^{n}$ elements is
unique up to isomorphism and is called the Galois field $\Gal{d^n}$.

We denote as $\mathbb{Z}_{d} [x]$ the ring of polynomials with
coefficients in $\mathbb{Z}_{d}$. If $P(x)$ is an irreducible
polynomial of degree $n$ (that is, one that cannot be factorized over
$\mathbb{Z}_{d}$),  the quotient space $\mathbb{Z}_{d}[X]/P(x)$
provides an adequate representation of $\Gal{d^n}$. Its elements can
be written as polynomials that are defined modulo the irreducible
polynomial $P(x)$. The multiplicative group of $\Gal{d^n}$ is cyclic
and its generator is called a primitive element of the field.

As a trivial example of a nonprime field, we consider the polynomial
$x^{2}+x+1=0$, which is irreducible over $\mathbb{Z}_{2}$.  If $\sigma$
is a root of this polynomial, the elements
$\{ 0, 1, \sigma, \sigma^{2} = \sigma+ 1 = \sigma^{-1} \} $ form the
finite field $\Gal{2^2}$ and $\sigma$ is a primitive element.

A basic map is the trace
\begin{equation}
  \label{deftr}
  \tr (\alpha ) = \alpha + \alpha^{d} + \ldots +
  \alpha^{d^{n-1}} \, .
\end{equation}
The image of the trace is always in the prime field
$\mathbb{Z}_d$ and satisfies
\begin{equation}
  \label{tracesum}
  \tr ( \alpha + \alpha^\prime ) =
  \tr ( \alpha ) + \tr ( \alpha^\prime ) \, .
\end{equation}
In terms of it we define an additive character as
\begin{equation}
  \label{Eq: addchardef}
  \chi (\alpha ) = \exp \left [ \frac{2 \pi i}{d}
    \tr ( \alpha ) \right] \, ,
\end{equation}
which possesses two important properties:
\begin{equation}
  \chi (\alpha + \alpha^\prime ) =
  \chi (\alpha ) \chi ( \alpha^\prime ) ,
  \qquad
  \sum_{\alpha^\prime \in
    \Gal{d^n}} \chi ( \alpha \alpha^\prime ) = d^n
  \delta_{0,\alpha} \, .
  \label{eq:addcharprop}
\end{equation}

Any finite field $\Gal{d^n}$ can be also considered as an
$n$-dimensional linear vector space over its prime field
  $\Gal{d}$. Given a basis $\{ \theta_{j} \}$, ($j = 1,\ldots, n$) in
this vector space, any field element can be represented as
\begin{equation}
  \label{mapnum}
  \alpha = \sum_{j=1}^{n} a_{j} \, \theta_{j} ,
\end{equation}
with $a_{j}\in \mathbb{Z}_{d}$. In this way, we map each element of
$\Gal{d^n}$ onto an ordered set of natural numbers
$\alpha \Leftrightarrow (a_{1}, \ldots , a_{n})$.

Two bases $\{ \theta_{1}, \ldots, \theta_{n} \} $ and
$\{ \theta_{1}^\prime, \ldots , \theta_{n}^\prime \} $ are dual when
\begin{equation}
  \tr ( \theta_{k} \theta_{l}^\prime ) =\delta_{k,l}.
\end{equation}
A basis that is dual to itself is called self-dual.  A self-dual basis
exists if and only if either $d$ is even or both $n$ and $d$ are odd.

There are several natural bases in $\Gal{d^n}$. One is the polynomial
basis, defined as
\begin{equation}
  \label{polynomial}
  \{1, \sigma, \sigma^{2}, \ldots, \sigma^{n-1} \} ,
\end{equation}
where $\sigma$ is a primitive element. An alternative is a
normal basis, constituted of
\begin{equation}
  \label{normal}
  \{\sigma, \sigma^{d}, \ldots, \sigma^{d^{n-1}} \}.
\end{equation}
The appropriate choice of basis depends on the specific problem at
hand. For example, in $\Gal{2^2}$ the elements
$\{ \sigma, \sigma^{2}\}$ are both roots of the irreducible
polynomial. The polynomial basis is $\{ 1, \sigma\} $ and its dual is
$\{ \sigma^{2}, 1 \}$, while the normal basis
$\{ \sigma, \sigma^{2} \} $ is self-dual.
  
\section{Derivation of equation for line operators}
\label{Sec: Lineop}

Here we present the derivation of our equation for the surviving
displacement operators.  We begin by considering the projectors for
the rays,
\begin{equation}
  \ket{\ell^{(\lambda)}_0}\bra{\ell^{(\lambda)}_0} = 
  \frac{1}{2^{mn}} \sum_{\alpha} D(\alpha, \lambda \alpha) = 
  \frac{1}{2^{mn}} \sum_{\alpha}  \Phi(\alpha, \lambda \alpha) 
  Z_\alpha X_{\lambda \alpha}\, .
\end{equation}
As mentioned in Sec.~\ref{sec:qpWig}, the projectors for the shifted lines can be
obtained by applying an appropriate displacement operator to induce a
transformation. Let us ignore for now the ray with infinite slope,
$\alpha = 0$.  Then for the rest of the rays, we can shift them
vertically by applying the displacement operators of the form
$D(0, \gamma)$:
\begin{eqnarray}
\ket{\ell^{(\lambda)}_\gamma}\bra{\ell^{(\lambda)}_\gamma}  &= & 
 \frac{1}{2^{mn}} \sum_{\alpha} D(0, \gamma) 
D(\alpha, \lambda \alpha)  D^\dag(0, \gamma) \nonumber \\
	&=& \frac{1}{2^{mn}} \sum_{\alpha} \Phi(\alpha, \lambda \alpha) 
  X_\gamma  Z_\alpha X_{\lambda \alpha} X_\gamma \, ,
\label{unfinished_line_with_intercept}
\end{eqnarray}
where we recall the convention that all the phases
$\Phi(0, \gamma) = 1$.

Here, we can make further use of the commutation relation in
Eq.~\eqref{commutation_relation}. We obtain
\begin{eqnarray}
\ket{\ell^{(\lambda)}_\gamma}\bra{\ell^{(\lambda)}_\gamma} &=& 
\frac{1}{2^{mn}} \sum_{\alpha} \Phi(\alpha, \lambda \alpha) 
\chi ( \gamma \alpha)  Z_\alpha X_{\lambda \alpha} \nonumber \\
&=& \frac{1}{2^{mn}} \sum_{\alpha} 
\chi ( \gamma \alpha) D \left(\alpha, \lambda \alpha \right).
\end{eqnarray}
It is then straightforward to see that the thick rays, which are
obtained by summing  over all intercepts $\gamma$ in coset
$\mathfrak{C_0}$, can be written as 
\begin{equation}
  \ket{{L}^{(\lambda)}_{\mathfrak{C}_0}} \bra{{L}^{(\lambda)}_{\mathfrak{C}_0}}
  = \frac{1}{2^{mn}} 
  \sum_{\lambda} 
\left [ 
\sum_{\gamma \in \mathfrak{C}_0}  \chi (\gamma \alpha) 
\right ] 
 D(\alpha, \lambda \alpha).
\end{equation}
Finally, we mention that for the infinite slope the analysis proceeds
in exactly the same way, but that the lines are translated by
displacement operators of the form $D(\gamma, 0)$ and
Eq. \eqref{commutation_relation} gives us $\chi (\gamma \beta)$ instead.

Only those operators which have a non-zero term in the sum will
contribute, thus we consider them as the effective displacement
operators in the coarse phase space.

\input{pra_v11.bbl}
\end{document}

%% file: pra_v11.bbl
%